\documentclass[reprint,amsmath,amssymb,aps,prl]{revtex4-1}
\usepackage{natbib}
\usepackage{makeidx}
\usepackage{url}
\usepackage[colorlinks=true,linkcolor=blue]{hyperref}

\usepackage{graphicx}
\usepackage{dcolumn}
\usepackage{bm}

\begin{document}

\preprint{APS/123-QED}

\title{Thermal excitations within and among mesospins in artificial spin ice}

\author{Bj\"orn Erik Skovdal}
\email{bjorn\_erik.skovdal@physics.uu.se}
\author{Samuel D. Sl\"{o}etjes}
\author{Merlin Pohlit}
\author{Henry Stopfel}
\author{Vassilios Kapaklis}
\author{Bj\"{o}rgvin Hj\"{o}rvarsson}

\affiliation{Department of Physics and Astronomy, Uppsala University, Box 516, 751 20 Uppsala, Sweden}

\begin{abstract}
We provide experimental and numerical evidence for thermal excitations within and among magnetic mesospins, forming artificial spin ice structures. At low temperatures, a decrease in magnetization and increase in susceptibility is observed with increasing temperature, interpreted as an onset of thermal fluctuations of the magnetic texture within the mesospins. At elevated temperatures a pronounced susceptibility peak is observed, related to thermally induced flipping of the mesospins and a collapse of the remanent state. The fluctuations, while occurring at distinct length- and energy-scales, are shown to be tunable by the interaction strength of the mesospins.
\end{abstract}

\maketitle

Mesoscopic spin systems, consisting of arrays of magnetostatically interacting ferromagnetic islands \cite{wang_artificial_2006}, are well suited to study phase transitions \cite{kapaklis_melting_2012, Anghinolfi:2015eu, sendetskyi_continuous_2019,levis2013thermal}, frustration \cite{Nisoli_2013,Nisoli:2017hg,gilbert2014emergent,drisko2017topological,morrison2013unhappy, ASI_Review_2020}, collective behavior \cite{Rougemaille_2019,ewerlin_magnetic_2013,jungfleisch2017high} as well as avalanches \cite{bingham2021experimental}. The \textit{mesospin} building blocks are often regarded as artificial ``atoms" without inner structure, where all the interactions, excitations and dynamics take place among and not within the mesospins. The use of mesospins allows for both magnetic imaging of individual elements \cite{wang_artificial_2006, Arnalds_2012_APL}, as well as adjusting many of the relevant parameters at will, such as interaction strength and temperature onset of fluctuations \cite{Andersson_2016_SciRep, Pohlit_PRB_2020}. For example, previous investigations include exploration of ordering and dynamics of mesospins, forming 1D and 2D lattices  \cite{ostman_ising-like_2018, arnalds2016new, Arnalds_XY, skovdal2021temperature}. 

Recent studies have begun to transcend this approach when including the effect of internal degrees of freedom (of the mesospin) on the order and ground state degeneracy of the arrangements \cite{skovdal2021temperature,Gliga_PRB_2015, Sloetjes_arXiv_2020}. For instance, internal vortex states in circular islands \cite{shinjo_magnetic_2000, Klaui_vortx_2003} can impact the interaction between islands and the global magnetic order, auguring first order magnetic phase transitions and tricritical behavior \cite{tricritical}. This exotic behavior is a consequence of the interplay between effects that take place on vastly different length scales, yet on comparable energy levels. 
Elongated Ising-like islands that make up artificial spin ice (ASI) also exhibit dynamic and quasi-static internal degrees of freedom in the form of curved magnetization at the \textit{edges} \cite{Sloetjes_arXiv_2020, phatak2011nanoscale}. At low temperatures, these S- and C-shaped texture states may appear as static textures that break the vertex symmetry in ASI, but at higher temperatures, dynamic edge modes have been predicted \cite{Sloetjes_arXiv_2020}, reducing the time averaged transverse magnetic moment within the islands to zero and restoring the vertex symmetry \cite{Gliga_PRB_2015}. Here we report on the interplay between the internal (within the mesospins) and external (among mesospins) excitations in ASI, and quantify their relative energy levels. We do this in both quasi-equilibrium and in out-of-equilibrium settings, using ac susceptibility and magnetization measurements respectively, comparing to results from stochastic micromagnetic simulations \cite{leliaert2017adaptively}. 

To this end, two samples were studied, having different mesospin interaction energies, tuned utilizing interaction modifiers (to be refered to as mASI) in the form of circular islands placed in the center of the ASI vertices \cite{ostman_interaction_2018}, while keeping the size and the distance between the Ising elements the same. Both samples were fabricated by post-patterning performed on a $\delta$-doped Pd(Fe) thin film, using electron beam lithography. The film, grown by dc magnetron sputtering, consisted of 40 nm palladium and 1.7 monolayers of iron with a 2 nm Palladium capping layer and a 1.5 nm vanadium seed layer on top of a magnesium oxide (MgO (001)) substrate as described in \citet{parnaste2007dimensionality}. The Ising-like islands had a length of 450 nm and a width of 150 nm with a pitch of 660 nm in both samples. The circular interaction modifiers at the ASI vertex sites, have a diameter of 130 nm. 

\begin{figure}[t!]
\begin{center}
\includegraphics[width=1\linewidth]{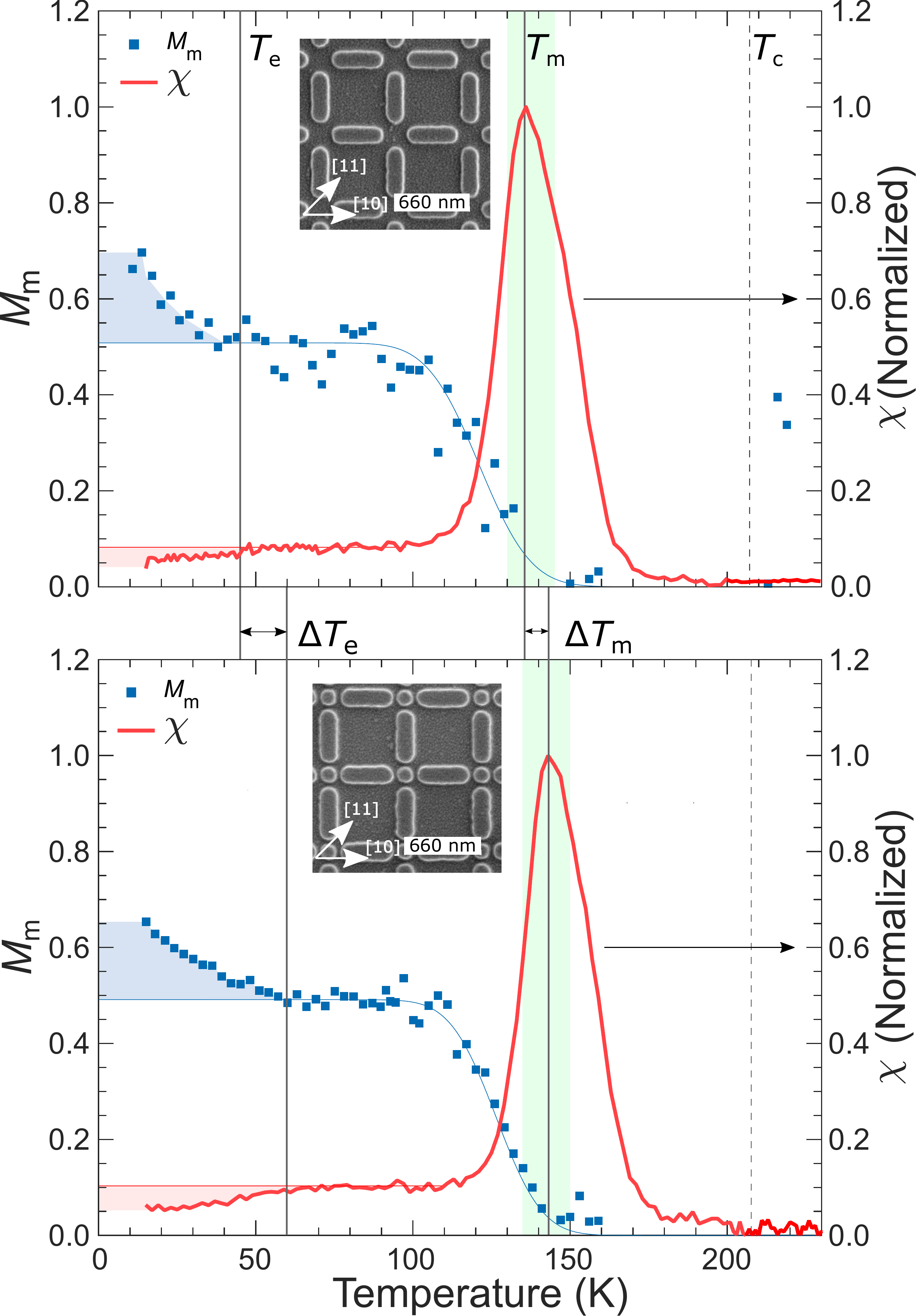}
\caption{Top panel: comparing $M_\mathrm{m}$ from ASI and for applied fields along the [10]-direction with the susceptibility, $\chi$, measured at 111 Hz. Below $T_\mathrm{e}$ the islands are frozen and exhibit static inner textures with a non-zero time-averaged transverse moment, $m_{\perp}$. Between $T_\mathrm{e}$ and $T_\mathrm{m}$, the islands exhibit texture dynamics of the edges. Above $T_\mathrm{m}$ the entire islands are fluctuating. The vertical lines are guides to the eye, indicating the relevant temperatures. The green shaded area indicates the span of $T_\mathrm{m}$ obtained from susceptibility measurements in the interval of 1 to 10000 Hz (from Fig. \ref{sus}). Bottom panel: Corresponding results from mASI. The values for $T_\mathrm{e}$ and $T_\mathrm{m}$ are shifted to higher temperatures, as indicated by $\Delta T_\mathrm{m}$ and $\Delta T_\mathrm{e}$ in-between the two panels. The insets show SEM images of the respective samples.}
\label{compare}
\end{center}
\end{figure}

Magnetization and ac susceptibility data were collected using magneto-optical Kerr effect (MOKE) in a longitudinal configuration, using $p$-polarized laser light with a wavelength of 660 nm. For the ac susceptibility measurements the field amplitude was chosen to be 0.02 mT and the frequency was varied between 1.1 and 3333~Hz. For the magnetization measurements the amplitude of the applied oscillatory field was 30~mT and the sweep frequency was 0.3~Hz. The samples were measured along both principal axes ([10] and [11], see insets of Fig. \ref{compare}) in the temperature range of $15 - 250$~K. In addition to the experimental characterization, we explored the dynamic behavior with micromagnetic simulations, using MuMax3 \cite{vansteenkiste_design_2014}. In these simulations, the temperature is emulated by a stochastic magnetic field as described by \citet{leliaert2017adaptively}. The micromagnetic constants used to describe the $\delta$-doped Pd(Fe) material are a saturation magnetization of $M_S = 3.5\times10^5$~A/m and an exchange stiffness of $A_{ex} = 3.36\times10^{-12}$~J/m, while the damping constant is taken to be $\alpha=0.02$. The simulations are run for 16 different temperatures between 0 and 400~K, for time intervals of 50~ns per temperature.

\begin{figure}[t!]
\begin{center}
\includegraphics[width=1\linewidth]{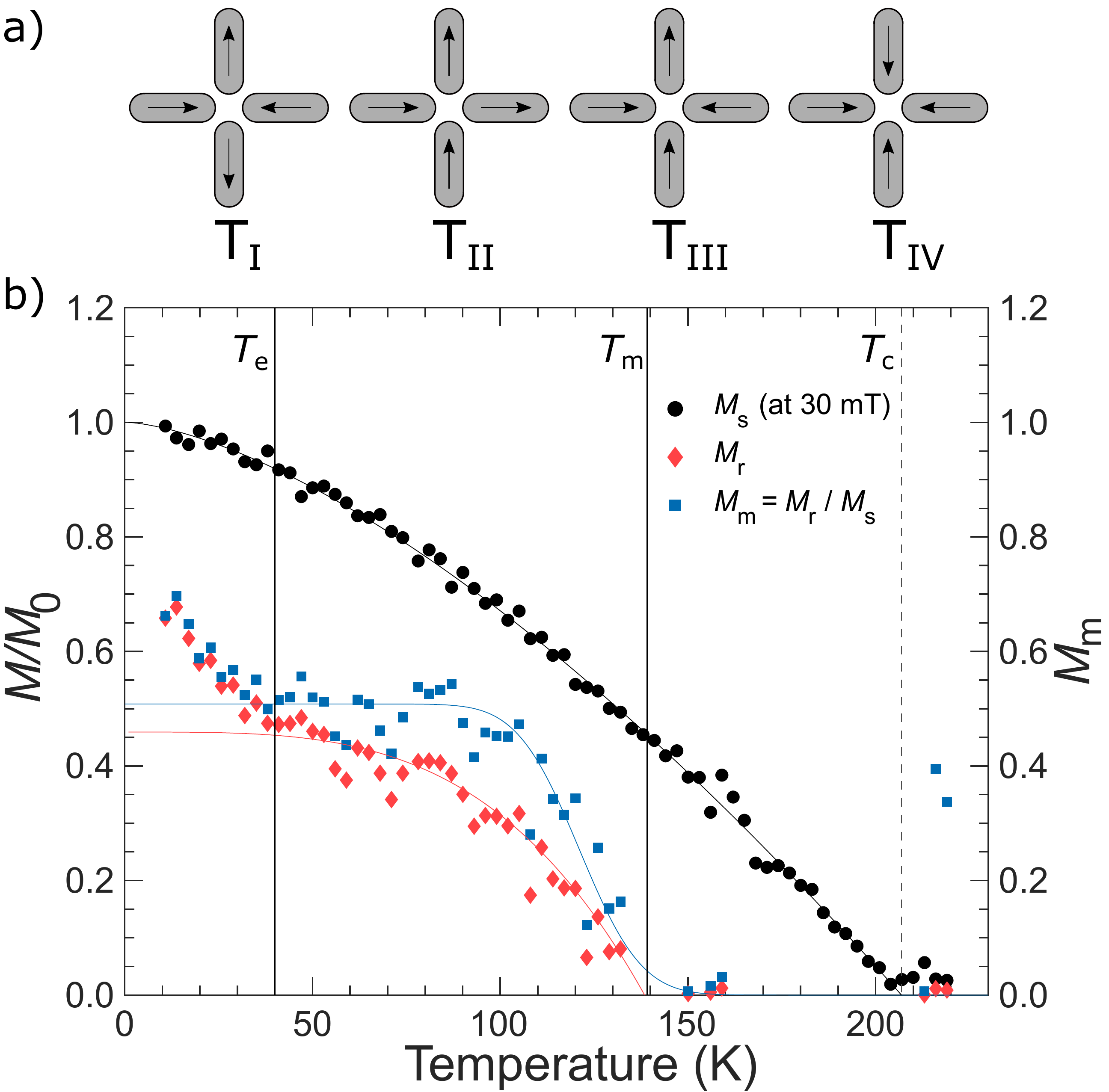}
\caption{ a) Schematic of the four types of vertices in ASI. b) Results from magnetization measurements where $M_\mathrm{s}$, $M_\mathrm{r}$ and $M_\mathrm{m}$ are plotted as a function of temperature with the field applied along the [10]-direction (see inset of Fig. \ref{compare}a). The vertical lines are guides to the eye, indicating the relevant fluctuation temperatures for the mesospin edges ($T_\mathrm{e}$), mesospins ($T_\mathrm{m}$) and the materials ordering temperature ($T_\mathrm{C}$).}
\label{MvsT}
\end{center}
\end{figure}

The main observables from the magnetization measurements, are the remanent magnetization obtained from the hysteresis loops, $M_\mathrm{r}$, and the saturation magnetization, $M_\mathrm{s}$, evaluated at $H~=~30$~mT. The temperature dependence of the magnetisation of the ASI array is illustrated in Fig. \ref{MvsT}.
The remanent magnetization, $M_\mathrm{r}$, is affected by thermal excitations on both the intra- and inter-mesospin length-scales. Separation of these contributions can be achieved by realizing that the $M_\mathrm{s}$ to a good approximation only depends on the intrinsic material properties, as a small applied field will easily quench all internal and external mesospin excitations, yet with a negligible impact on the temperature dependent material magnetization. The quantity $M_\mathrm{m} = M_\mathrm{r}/M_\mathrm{s}$ is therefore taken as a measure of the contribution from the mesospins, which are thereby separated from the fluctuations at the atomic scale. Furthermore, the temperature at which $M_\mathrm{s}$ approaches zero is an approximate measure of the ordering temperature, $T_\mathrm{c}$, of the material.

We proceed by comparing the results from the magnetization of the mesospins ($M_\mathrm{m}$) and the magnetic susceptibility obtained for the ASI and mASI samples, as shown in Fig. \ref{compare}. Starting form high temperatures, we notice no peak or other feature in the susceptibility data at $T_\mathrm{c}$ of the material (marked by the dashed line). Instead, a peak appears at the temperature where $M_\mathrm{m}$ approaches zero, which we label $T_\mathrm{m}$. Since the mesospins should carry a moment below $T_\mathrm{c}$, the fact that $M_\mathrm{m} = 0$ between $T_\mathrm{m}$ and $T_\mathrm{c}$ suggests that the magnetization vanishes as a consequence of fluctuations of the mesospins, with a characteristic time-scale shorter than that of the measurement \cite{Pohlit_PRB_2020}. The susceptibility peak is therefore related to mesospin fluctuations rather than to fluctuations at the atomic scale. Here, the value of $T_\mathrm{m}$ is taken to be the temperature at which the average relaxation time is equal to the time window of the measurements.

Lowering the temperature further, we find a plateau between 45 and 100~K, in both the susceptibility and the magnetization curves. The height of this plateau is at $M_\mathrm{m} \approx 0.5$ for measurements along the [10]-direction and $M_\mathrm{m} \approx 0.7 \approx \sqrt2$ for the [11]-direction, as has been previously reported by \citet{kapaklis_melting_2012}. These values of $M_\mathrm{m}$ agree with the expected value of an ASI lattice comprised of frozen Ising-like mesospins, with a zero net transverse magnetization in the islands. Turning to the susceptibility, the plateau means that some part of the system does respond to the applied ac field. Knowing that the mesospins are frozen at these temperatures with respect to magnetization flipping, the observed susceptibility must originate from internal magnetic excitations of the elements. We attribute this finite susceptibility to be related to dynamics of the magnetization texture at the mesospin edges, since the continuous $\delta$-doped Pd(Fe) thin film exhibits zero susceptibility in this temperature regime (not shown here). Thus, the edges fluctuate at these temperatures, resulting in a zero time averaged transverse moment in the islands.

Turning our attention to the lowest temperatures measured, we find that $M_\mathrm{m}$ begins to increase below $40 \pm 3$ K while the susceptibility decreases in the same temperature range ($\leq 45 \pm 3$ K). An increase in $M_\mathrm{m}$ beyond 0.5 requires the presence of an internal transverse magnetic component of the islands. Therefore, this temperature labeled $T_\mathrm{e}$, marks the onset of textures stabilizing that exhibit a non-zero time average magnetization in the direction of the applied field. This effect can therefore be understood as a consequence of a slow low-temperature relaxation of the edges, i.e. the edges remain static within the time-scale of the measurement. This interpretation is supported by the observed decrease in susceptibility with decreasing temperature. 

\begin{figure}[t!]
\begin{center}
\includegraphics[width=1\linewidth]{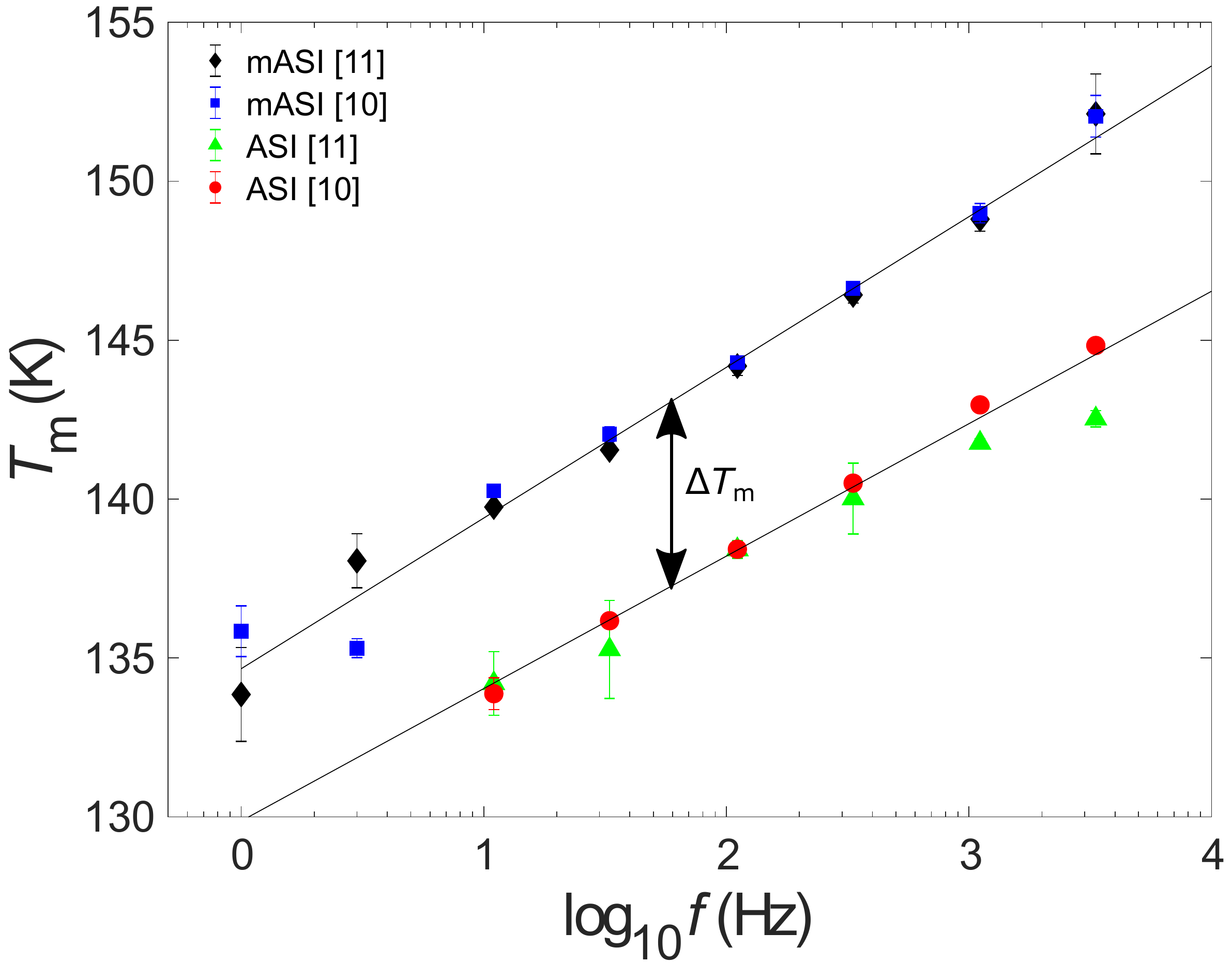}
\caption{Susceptibility peak position as a function of sweep frequency for both samples and ac fields applied along the [10] and [11]-direction. The difference, $\Delta T_\mathrm{m}$, between the ASI and mASI is determined to be $6 \pm 1$ K.}
\label{sus}
\end{center}
\end{figure}

It is important to mention that the susceptibility measurements probe equilibrium dynamics, while the magnetization measurements probe field-``dressed" metastable states. This gives rise to subtle differences in the results depending on which measurement protocol is used. Firstly, when measuring along [11], the initial state (after removing the field) is a uniform state of exclusively $\mathrm{T_{II}}$ vertices, while measuring along [10] leads to a mixture of $\mathrm{T_{II}}$ and $\mathrm{T_{III}}$ vertices. As the $\mathrm{T_{III}}$ vertex is a higher energy excitation than $\mathrm{T_{II}}$, we expect somewhat different relaxation depending on the orientation of the measurement, and there is no reason to choose the [10] magnetization data over the [11] for comparison. Secondly, the position of the susceptibility peak is frequency dependent as seen in Fig. \ref{sus} and reported by \citet{Pohlit_PRB_2020}, which implies that the position of $T_\mathrm{m}$ as obtained from the magnetization measurements also depends on the sweep frequency. To verify that the observed equivalence of $T_\mathrm{m}$ in the susceptibility and magnetization data (similar for $T_\mathrm{e}$) is not purely coincidental, one would have to unify the time-scales of the different measurement techniques. This proves to be challenging as the amplitudes of the field sweeps also differ, and we therefore propose an alternative route to comparing the outcomes of the measurement schemes.

Using the interaction modifier in the mASI array, it is possible to perform a relative comparison between both techniques. In Fig.~\ref{sus}, we have plotted $T_\mathrm{m}$ as a function of sweep frequency for both ASI and mASI along both [10] and [11]. From this graph, we notice that although $T_\mathrm{m}$ varies with sweep frequency, it only changes by about 15 K when changing the time-scale by four orders of magnitude. Thus, the overlap in $T_\mathrm{m}$ should be maintained even for large time-scale discrepancies between magnetization and susceptibility measurements. Moreover, we notice that there is no orientation dependence in the susceptibility measurements, which is expected as these measurements probe the response of the ground state, while the magnetisation results represent dressed states. Lastly, we find that the difference between the samples with and without an interaction modifier, indicated by $\Delta T_\mathrm{m}$, is seemingly independent on the sweep frequency, and has a value of $6 \pm 1$ K. We can use this fact and compare $\Delta T_\mathrm{m}$ from the susceptibility measurements with $\Delta T_\mathrm{m}$ from the magnetization measurements for both [10] and [11]-directions.

While $\Delta T_\mathrm{m}$ from the susceptibility measurements is robust and precise, accurately quantifying the individual values of $T_\mathrm{m}$ from the magnetization measurements is challenging as they are highly sensitive to the fitting procedure, in particular to the number of points included in the fits. Therefore, we chose to use the centroid of the error function fit to $M_\mathrm{m}$ as a measure of $T_\mathrm{m}$. This parameter is less sensitive to the fitting procedure, and yields a value that is suitable for comparison between the two samples. It will, however, include an offset with respect to the true $T_\mathrm{m}$ which we assume is constant for all measurements. Since the magnetization results differ depending on measurement orientation, we will compare [10] and [11]-direction separately. The difference in centroid temperature between ASI and mASI is $5.4 \pm 0.9$ K for [10] and $4 \pm 0.6$ K for [11], which agrees well with the $6 \pm 1$ K obtained from the susceptibility results. All of these findings hint towards the overlap in $T_\mathrm{m}$ between $M_\mathrm{m}$ and $\chi$ not being accidental.
The same argument can be used when analyzing the low temperature behavior. Here, $T_\mathrm{e}$ as obtained from the susceptibility data is $45 \pm 3$ and $65 \pm 3$~K for ASI and mASI, respectively and the corresponding figures obtained from the magnetization data are $40 \pm 3$ and $60 \pm 3$~K (see Fig. \ref{compare}). The low temperature increase in $M_\mathrm{m}$, and decrease in $\chi$ suggests a decrease in dynamics on these time-scales. We therefore interpret $T_\mathrm{e}$ as the ``melting" temperature of the edge textures, predicted by \citet{Gliga_PRB_2015} and \citet{Sloetjes_arXiv_2020}.

Consequently, we performed micromagnetic simulations to explore the feasibility of the hypothesis concerning the dynamics of the inner magnetic texture.
\begin{figure}[t!]
\begin{center}
\includegraphics[width=\linewidth]{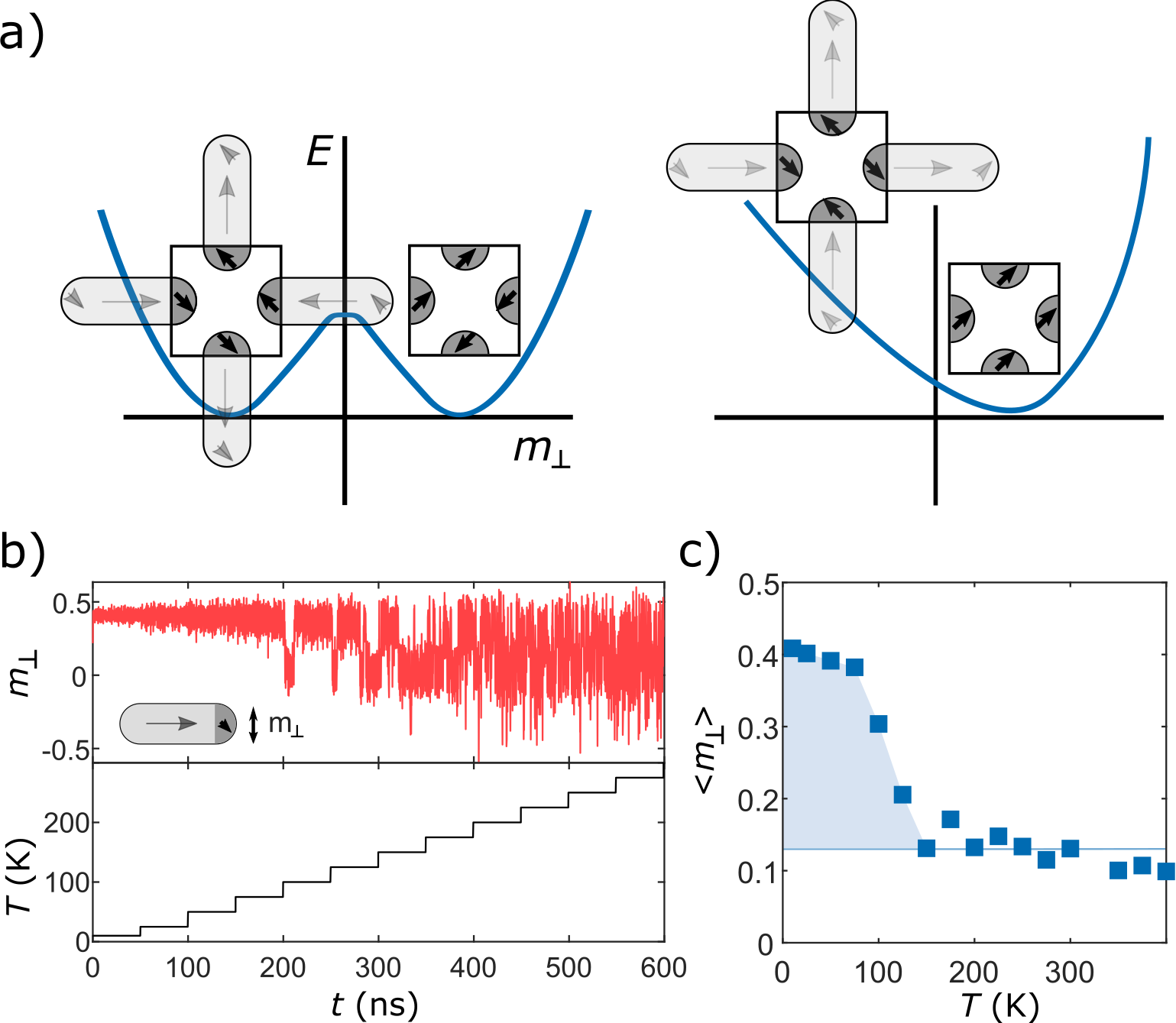}
\caption{(a) Schematics of the energy landscape associated with the edge magnetization in the $\mathrm{T_I}$ (left) and $\mathrm{T_{II}}$ vertex (right). (b) The temporal evolution of perpendicular edge magnetization fluctuations at different temperatures (indicated in the lower panel). (c) Time averaged magnetization values at different temperatures.}
\label{sims}
\end{center}
\end{figure} 
Thermal edge fluctuations in $\mathrm{T_I}$ and $\mathrm{T_{II}}$ vertices are different in nature due to differences in the energy landscape morphology (see Fig. \ref{sims}a). A $\mathrm{T_I}$ vertex features a double-well potential, whereas $\mathrm{T_{II}}$ has an asymmetric single-well potential. The signal measured in the experiments is a result of a weighted sum of contributions from the different vertex types, therefore we simulated this system with equal weight from $\mathrm{T_I}$ and $\mathrm{T_{II}}$ vertices. The resulting timetraces of the fluctuations are shown in the upper panel of Fig. \ref{sims}b, and the corresponding temperatures can be seen in the lower panel. Here, the perpendicular magnetization is defined as the total  perpendicular magnetization in one edge (defined by a half circle), normalized by the saturation magnetization, i.e. $m_\perp = M_\perp^{edge}/M_S$. A signature of the double well potential is the telegraph noise in the magnetization, which can be observed in the signal, whereas the fluctuations associated with the $\mathrm{T_{II}}$ vertex provide a bias to positive values of $m_\perp$. The time-average magnetization, $\langle m_\perp \rangle$, shown in Fig. \ref{sims}c, converges to a value of $m_\perp = 0.1$ at high temperatures. As the temperature is lowered, the fluctuations start to freeze out at T = 150 K, reaching a value of $m_\perp = 0.4$ below T = 50 K. This finite value of $m_\perp$ is mostly due to the freezing out of the fluctuations in the asymmetric potential associated with the $\mathrm{T_{II}}$ vertices. The upturn of the magnetization at low temperatures is in qualitative agreement with experiments, and the quantitative discrepancy in $T_e$ can be ascribed to the vastly different time-scales between simulations and experiments. The lack of fluctuations at low temperatures also explains the decrease in the ac susceptibility signal seen in the experiments.
We provide experimental evidence of edge melting as well as a mesospin melting in artificial spin ice structures. A change in the interactions has a measurable impact on the onset of these fluctuations. The temperature at which the mesospins start fluctuating is approximately a factor 3--3.5 times higher than the temperature at which the edge fluctuations are obtained. While the edge fluctuations have been shown to impact resonant spectra features of ASI lattices \cite{Gliga_PRB_2015}, their effect on the ordering has not been demonstrated before. 
For the case presented here, the energy levels of the fluctuations are well separated, but a question may be raised as to how a degeneracy of the levels would affect the dynamics of the mesospins.

\section*{Acknowledgments}
B.H. and V.K. acknowledge financial support from the National Research Council (VR) (Project No. 2019-03581 and {2019-05379}). We acknowledge Myfab Uppsala for providing facilities and experimental support. Myfab is funded by the Swedish Research Council (2019-00207) as a national research infrastructure. This research used resources of the Center for Functional Nanomaterials (CFN), which is a U.S. Department of Energy Office of Science User Facility, at Brookhaven National Laboratory under Contract No. DE-SC0012704.

\section*{Author Declarations}

\subsection*{Data availability}
The data that support the findings are available from the corresponding authors upon reasonable request.

\subsection*{Conflict of Interest}
The authors have no conflicts to disclose.


\begin{thebibliography}{32}%
\makeatletter
\providecommand \@ifxundefined [1]{%
 \@ifx{#1\undefined}
}%
\providecommand \@ifnum [1]{%
 \ifnum #1\expandafter \@firstoftwo
 \else \expandafter \@secondoftwo
 \fi
}%
\providecommand \@ifx [1]{%
 \ifx #1\expandafter \@firstoftwo
 \else \expandafter \@secondoftwo
 \fi
}%
\providecommand \natexlab [1]{#1}%
\providecommand \enquote  [1]{``#1''}%
\providecommand \bibnamefont  [1]{#1}%
\providecommand \bibfnamefont [1]{#1}%
\providecommand \citenamefont [1]{#1}%
\providecommand \href@noop [0]{\@secondoftwo}%
\providecommand \href [0]{\begingroup \@sanitize@url \@href}%
\providecommand \@href[1]{\@@startlink{#1}\@@href}%
\providecommand \@@href[1]{\endgroup#1\@@endlink}%
\providecommand \@sanitize@url [0]{\catcode `\\12\catcode `\$12\catcode
  `\&12\catcode `\#12\catcode `\^12\catcode `\_12\catcode `\%12\relax}%
\providecommand \@@startlink[1]{}%
\providecommand \@@endlink[0]{}%
\providecommand \url  [0]{\begingroup\@sanitize@url \@url }%
\providecommand \@url [1]{\endgroup\@href {#1}{\urlprefix }}%
\providecommand \urlprefix  [0]{URL }%
\providecommand \Eprint [0]{\href }%
\providecommand \doibase [0]{http://dx.doi.org/}%
\providecommand \selectlanguage [0]{\@gobble}%
\providecommand \bibinfo  [0]{\@secondoftwo}%
\providecommand \bibfield  [0]{\@secondoftwo}%
\providecommand \translation [1]{[#1]}%
\providecommand \BibitemOpen [0]{}%
\providecommand \bibitemStop [0]{}%
\providecommand \bibitemNoStop [0]{.\EOS\space}%
\providecommand \EOS [0]{\spacefactor3000\relax}%
\providecommand \BibitemShut  [1]{\csname bibitem#1\endcsname}%
\let\auto@bib@innerbib\@empty
\bibitem [{\citenamefont {Wang}\ \emph {et~al.}(2006)\citenamefont {Wang},
  \citenamefont {Nisoli}, \citenamefont {Freitas}, \citenamefont {Li},
  \citenamefont {McConville}, \citenamefont {Cooley}, \citenamefont {Lund},
  \citenamefont {Samarth}, \citenamefont {Leighton}, \citenamefont {Crespi},\
  and\ \citenamefont {Schiffer}}]{wang_artificial_2006}%
  \BibitemOpen
  \bibfield  {author} {\bibinfo {author} {\bibfnamefont {R.~F.}\ \bibnamefont
  {Wang}}, \bibinfo {author} {\bibfnamefont {C.}~\bibnamefont {Nisoli}},
  \bibinfo {author} {\bibfnamefont {R.~S.}\ \bibnamefont {Freitas}}, \bibinfo
  {author} {\bibfnamefont {J.}~\bibnamefont {Li}}, \bibinfo {author}
  {\bibfnamefont {W.}~\bibnamefont {McConville}}, \bibinfo {author}
  {\bibfnamefont {B.~J.}\ \bibnamefont {Cooley}}, \bibinfo {author}
  {\bibfnamefont {M.~S.}\ \bibnamefont {Lund}}, \bibinfo {author}
  {\bibfnamefont {N.}~\bibnamefont {Samarth}}, \bibinfo {author} {\bibfnamefont
  {C.}~\bibnamefont {Leighton}}, \bibinfo {author} {\bibfnamefont {V.~H.}\
  \bibnamefont {Crespi}}, \ and\ \bibinfo {author} {\bibfnamefont
  {P.}~\bibnamefont {Schiffer}},\ }\href {\doibase 10.1038/nature04447}
  {\bibfield  {journal} {\bibinfo  {journal} {Nature}\ }\textbf {\bibinfo
  {volume} {439}},\ \bibinfo {pages} {303} (\bibinfo {year}
  {2006})}\BibitemShut {NoStop}%
\bibitem [{\citenamefont {Kapaklis}\ \emph {et~al.}(2012)\citenamefont
  {Kapaklis}, \citenamefont {Arnalds}, \citenamefont {Harman-Clarke},
  \citenamefont {Papaioannou}, \citenamefont {Karimipour}, \citenamefont
  {Korelis}, \citenamefont {Taroni}, \citenamefont {Holdsworth}, \citenamefont
  {Bramwell},\ and\ \citenamefont {Hj{\"o}rvarsson}}]{kapaklis_melting_2012}%
  \BibitemOpen
  \bibfield  {author} {\bibinfo {author} {\bibfnamefont {V.}~\bibnamefont
  {Kapaklis}}, \bibinfo {author} {\bibfnamefont {U.~B.}\ \bibnamefont
  {Arnalds}}, \bibinfo {author} {\bibfnamefont {A.}~\bibnamefont
  {Harman-Clarke}}, \bibinfo {author} {\bibfnamefont {E.~T.}\ \bibnamefont
  {Papaioannou}}, \bibinfo {author} {\bibfnamefont {M.}~\bibnamefont
  {Karimipour}}, \bibinfo {author} {\bibfnamefont {P.}~\bibnamefont {Korelis}},
  \bibinfo {author} {\bibfnamefont {A.}~\bibnamefont {Taroni}}, \bibinfo
  {author} {\bibfnamefont {P.~C.~W.}\ \bibnamefont {Holdsworth}}, \bibinfo
  {author} {\bibfnamefont {S.~T.}\ \bibnamefont {Bramwell}}, \ and\ \bibinfo
  {author} {\bibfnamefont {B.}~\bibnamefont {Hj{\"o}rvarsson}},\ }\href
  {\doibase 10.1088/1367-2630/14/3/035009} {\bibfield  {journal} {\bibinfo
  {journal} {New Journal of Physics}\ }\textbf {\bibinfo {volume} {14}},\
  \bibinfo {pages} {035009} (\bibinfo {year} {2012})}\BibitemShut {NoStop}%
\bibitem [{\citenamefont {Anghinolfi}\ \emph {et~al.}(2015)\citenamefont
  {Anghinolfi}, \citenamefont {Luetkens}, \citenamefont {Perron}, \citenamefont
  {Flokstra}, \citenamefont {Sendetskyi}, \citenamefont {Suter}, \citenamefont
  {Prokscha}, \citenamefont {Derlet}, \citenamefont {Lee},\ and\ \citenamefont
  {Heyderman}}]{Anghinolfi:2015eu}%
  \BibitemOpen
  \bibfield  {author} {\bibinfo {author} {\bibfnamefont {L.}~\bibnamefont
  {Anghinolfi}}, \bibinfo {author} {\bibfnamefont {H.}~\bibnamefont
  {Luetkens}}, \bibinfo {author} {\bibfnamefont {J.}~\bibnamefont {Perron}},
  \bibinfo {author} {\bibfnamefont {M.~G.}\ \bibnamefont {Flokstra}}, \bibinfo
  {author} {\bibfnamefont {O.}~\bibnamefont {Sendetskyi}}, \bibinfo {author}
  {\bibfnamefont {A.}~\bibnamefont {Suter}}, \bibinfo {author} {\bibfnamefont
  {T.}~\bibnamefont {Prokscha}}, \bibinfo {author} {\bibfnamefont {P.~M.}\
  \bibnamefont {Derlet}}, \bibinfo {author} {\bibfnamefont {S.~L.}\
  \bibnamefont {Lee}}, \ and\ \bibinfo {author} {\bibfnamefont {L.~J.}\
  \bibnamefont {Heyderman}},\ }\href {\doibase 10.1038/ncomms9278} {\bibfield
  {journal} {\bibinfo  {journal} {Nature Communications}\ }\textbf {\bibinfo
  {volume} {6}},\ \bibinfo {pages} {8278} (\bibinfo {year} {2015})}\BibitemShut
  {NoStop}%
\bibitem [{\citenamefont {Sendetskyi}\ \emph {et~al.}(2019)\citenamefont
  {Sendetskyi}, \citenamefont {Scagnoli}, \citenamefont {Leo}, \citenamefont
  {Anghinolfi}, \citenamefont {Alberca}, \citenamefont {L{\"u}ning},
  \citenamefont {Staub}, \citenamefont {Derlet},\ and\ \citenamefont
  {Heyderman}}]{sendetskyi_continuous_2019}%
  \BibitemOpen
  \bibfield  {author} {\bibinfo {author} {\bibfnamefont {O.}~\bibnamefont
  {Sendetskyi}}, \bibinfo {author} {\bibfnamefont {V.}~\bibnamefont
  {Scagnoli}}, \bibinfo {author} {\bibfnamefont {N.}~\bibnamefont {Leo}},
  \bibinfo {author} {\bibfnamefont {L.}~\bibnamefont {Anghinolfi}}, \bibinfo
  {author} {\bibfnamefont {A.}~\bibnamefont {Alberca}}, \bibinfo {author}
  {\bibfnamefont {J.}~\bibnamefont {L{\"u}ning}}, \bibinfo {author}
  {\bibfnamefont {U.}~\bibnamefont {Staub}}, \bibinfo {author} {\bibfnamefont
  {P.~M.}\ \bibnamefont {Derlet}}, \ and\ \bibinfo {author} {\bibfnamefont
  {L.~J.}\ \bibnamefont {Heyderman}},\ }\href {\doibase
  10.1103/PhysRevB.99.214430} {\bibfield  {journal} {\bibinfo  {journal}
  {Physical Review B}\ }\textbf {\bibinfo {volume} {99}},\ \bibinfo {pages}
  {214430} (\bibinfo {year} {2019})}\BibitemShut {NoStop}%
\bibitem [{\citenamefont {Levis}\ \emph {et~al.}(2013)\citenamefont {Levis},
  \citenamefont {Cugliandolo}, \citenamefont {Foini},\ and\ \citenamefont
  {Tarzia}}]{levis2013thermal}%
  \BibitemOpen
  \bibfield  {author} {\bibinfo {author} {\bibfnamefont {D.}~\bibnamefont
  {Levis}}, \bibinfo {author} {\bibfnamefont {L.~F.}\ \bibnamefont
  {Cugliandolo}}, \bibinfo {author} {\bibfnamefont {L.}~\bibnamefont {Foini}},
  \ and\ \bibinfo {author} {\bibfnamefont {M.}~\bibnamefont {Tarzia}},\
  }\href@noop {} {\bibfield  {journal} {\bibinfo  {journal} {Physical review
  letters}\ }\textbf {\bibinfo {volume} {110}},\ \bibinfo {pages} {207206}
  (\bibinfo {year} {2013})}\BibitemShut {NoStop}%
\bibitem [{\citenamefont {Nisoli}\ \emph {et~al.}(2013)\citenamefont {Nisoli},
  \citenamefont {Moessner},\ and\ \citenamefont {Schiffer}}]{Nisoli_2013}%
  \BibitemOpen
  \bibfield  {author} {\bibinfo {author} {\bibfnamefont {C.}~\bibnamefont
  {Nisoli}}, \bibinfo {author} {\bibfnamefont {R.}~\bibnamefont {Moessner}}, \
  and\ \bibinfo {author} {\bibfnamefont {P.}~\bibnamefont {Schiffer}},\ }\href
  {\doibase 10.1103/RevModPhys.85.1473} {\bibfield  {journal} {\bibinfo
  {journal} {Rev. Mod. Phys.}\ }\textbf {\bibinfo {volume} {85}},\ \bibinfo
  {pages} {1473} (\bibinfo {year} {2013})}\BibitemShut {NoStop}%
\bibitem [{\citenamefont {Nisoli}\ \emph {et~al.}(2017)\citenamefont {Nisoli},
  \citenamefont {Kapaklis},\ and\ \citenamefont {Schiffer}}]{Nisoli:2017hg}%
  \BibitemOpen
  \bibfield  {author} {\bibinfo {author} {\bibfnamefont {C.}~\bibnamefont
  {Nisoli}}, \bibinfo {author} {\bibfnamefont {V.}~\bibnamefont {Kapaklis}}, \
  and\ \bibinfo {author} {\bibfnamefont {P.}~\bibnamefont {Schiffer}},\ }\href
  {\doibase 10.1038/nphys4059} {\bibfield  {journal} {\bibinfo  {journal}
  {Nature Physics}\ }\textbf {\bibinfo {volume} {13}},\ \bibinfo {pages} {200}
  (\bibinfo {year} {2017})}\BibitemShut {NoStop}%
\bibitem [{\citenamefont {Gilbert}\ \emph {et~al.}(2014)\citenamefont
  {Gilbert}, \citenamefont {Chern}, \citenamefont {Zhang}, \citenamefont
  {O’Brien}, \citenamefont {Fore}, \citenamefont {Nisoli},\ and\
  \citenamefont {Schiffer}}]{gilbert2014emergent}%
  \BibitemOpen
  \bibfield  {author} {\bibinfo {author} {\bibfnamefont {I.}~\bibnamefont
  {Gilbert}}, \bibinfo {author} {\bibfnamefont {G.-W.}\ \bibnamefont {Chern}},
  \bibinfo {author} {\bibfnamefont {S.}~\bibnamefont {Zhang}}, \bibinfo
  {author} {\bibfnamefont {L.}~\bibnamefont {O’Brien}}, \bibinfo {author}
  {\bibfnamefont {B.}~\bibnamefont {Fore}}, \bibinfo {author} {\bibfnamefont
  {C.}~\bibnamefont {Nisoli}}, \ and\ \bibinfo {author} {\bibfnamefont
  {P.}~\bibnamefont {Schiffer}},\ }\href {\doibase 10.1038/nphys3037}
  {\bibfield  {journal} {\bibinfo  {journal} {Nature Physics}\ }\textbf
  {\bibinfo {volume} {10}},\ \bibinfo {pages} {670} (\bibinfo {year}
  {2014})}\BibitemShut {NoStop}%
\bibitem [{\citenamefont {Drisko}\ \emph {et~al.}(2017)\citenamefont {Drisko},
  \citenamefont {Marsh},\ and\ \citenamefont
  {Cumings}}]{drisko2017topological}%
  \BibitemOpen
  \bibfield  {author} {\bibinfo {author} {\bibfnamefont {J.}~\bibnamefont
  {Drisko}}, \bibinfo {author} {\bibfnamefont {T.}~\bibnamefont {Marsh}}, \
  and\ \bibinfo {author} {\bibfnamefont {J.}~\bibnamefont {Cumings}},\ }\href
  {\doibase 10.1038/ncomms14009} {\bibfield  {journal} {\bibinfo  {journal}
  {Nature communications}\ }\textbf {\bibinfo {volume} {8}},\ \bibinfo {pages}
  {1} (\bibinfo {year} {2017})}\BibitemShut {NoStop}%
\bibitem [{\citenamefont {Morrison}\ \emph {et~al.}(2013)\citenamefont
  {Morrison}, \citenamefont {Nelson},\ and\ \citenamefont
  {Nisoli}}]{morrison2013unhappy}%
  \BibitemOpen
  \bibfield  {author} {\bibinfo {author} {\bibfnamefont {M.~J.}\ \bibnamefont
  {Morrison}}, \bibinfo {author} {\bibfnamefont {T.~R.}\ \bibnamefont
  {Nelson}}, \ and\ \bibinfo {author} {\bibfnamefont {C.}~\bibnamefont
  {Nisoli}},\ }\href {\doibase 10.1088/1367-2630/15/4/045009} {\bibfield
  {journal} {\bibinfo  {journal} {New Journal of Physics}\ }\textbf {\bibinfo
  {volume} {15}},\ \bibinfo {pages} {045009} (\bibinfo {year}
  {2013})}\BibitemShut {NoStop}%
\bibitem [{\citenamefont {Skjærvø}\ \emph {et~al.}(2020)\citenamefont
  {Skjærvø}, \citenamefont {Marrows}, \citenamefont {Stamps},\ and\
  \citenamefont {Heyderman}}]{ASI_Review_2020}%
  \BibitemOpen
  \bibfield  {author} {\bibinfo {author} {\bibfnamefont {S.~H.}\ \bibnamefont
  {Skjærvø}}, \bibinfo {author} {\bibfnamefont {C.~H.}\ \bibnamefont
  {Marrows}}, \bibinfo {author} {\bibfnamefont {R.~L.}\ \bibnamefont {Stamps}},
  \ and\ \bibinfo {author} {\bibfnamefont {L.~J.}\ \bibnamefont {Heyderman}},\
  }\href {\doibase 10.1038/s42254-019-0118-3} {\bibfield  {journal} {\bibinfo
  {journal} {Nature Reviews Physics}\ }\textbf {\bibinfo {volume} {2}},\
  \bibinfo {pages} {13} (\bibinfo {year} {2020})}\BibitemShut {NoStop}%
\bibitem [{\citenamefont {Rougemaille}\ and\ \citenamefont
  {Canals}(2019)}]{Rougemaille_2019}%
  \BibitemOpen
  \bibfield  {author} {\bibinfo {author} {\bibfnamefont {N.}~\bibnamefont
  {Rougemaille}}\ and\ \bibinfo {author} {\bibfnamefont {B.}~\bibnamefont
  {Canals}},\ }\href {\doibase 10.1140/ep jb/e2018-90346-7} {\bibfield
  {journal} {\bibinfo  {journal} {Eur. Phys. J. B}\ }\textbf {\bibinfo {volume}
  {92}},\ \bibinfo {pages} {62} (\bibinfo {year} {2019})}\BibitemShut {NoStop}%
\bibitem [{\citenamefont {Ewerlin}\ \emph {et~al.}(2013)\citenamefont
  {Ewerlin}, \citenamefont {Demirbas}, \citenamefont {Br{\"u}ssing},
  \citenamefont {Petracic}, \citenamefont {{\"U}nal}, \citenamefont {Valencia},
  \citenamefont {Kronast},\ and\ \citenamefont
  {Zabel}}]{ewerlin_magnetic_2013}%
  \BibitemOpen
  \bibfield  {author} {\bibinfo {author} {\bibfnamefont {M.}~\bibnamefont
  {Ewerlin}}, \bibinfo {author} {\bibfnamefont {D.}~\bibnamefont {Demirbas}},
  \bibinfo {author} {\bibfnamefont {F.}~\bibnamefont {Br{\"u}ssing}}, \bibinfo
  {author} {\bibfnamefont {O.}~\bibnamefont {Petracic}}, \bibinfo {author}
  {\bibfnamefont {A.~A.}\ \bibnamefont {{\"U}nal}}, \bibinfo {author}
  {\bibfnamefont {S.}~\bibnamefont {Valencia}}, \bibinfo {author}
  {\bibfnamefont {F.}~\bibnamefont {Kronast}}, \ and\ \bibinfo {author}
  {\bibfnamefont {H.}~\bibnamefont {Zabel}},\ }\href {\doibase
  10.1103/PhysRevLett.110.177209} {\bibfield  {journal} {\bibinfo  {journal}
  {Physical Review Letters}\ }\textbf {\bibinfo {volume} {110}},\ \bibinfo
  {pages} {177209} (\bibinfo {year} {2013})}\BibitemShut {NoStop}%
\bibitem [{\citenamefont {Jungfleisch}\ \emph {et~al.}(2017)\citenamefont
  {Jungfleisch}, \citenamefont {Sklenar}, \citenamefont {Ding}, \citenamefont
  {Park}, \citenamefont {Pearson}, \citenamefont {Novosad}, \citenamefont
  {Schiffer},\ and\ \citenamefont {Hoffmann}}]{jungfleisch2017high}%
  \BibitemOpen
  \bibfield  {author} {\bibinfo {author} {\bibfnamefont {M.~B.}\ \bibnamefont
  {Jungfleisch}}, \bibinfo {author} {\bibfnamefont {J.}~\bibnamefont
  {Sklenar}}, \bibinfo {author} {\bibfnamefont {J.}~\bibnamefont {Ding}},
  \bibinfo {author} {\bibfnamefont {J.}~\bibnamefont {Park}}, \bibinfo {author}
  {\bibfnamefont {J.~E.}\ \bibnamefont {Pearson}}, \bibinfo {author}
  {\bibfnamefont {V.}~\bibnamefont {Novosad}}, \bibinfo {author} {\bibfnamefont
  {P.}~\bibnamefont {Schiffer}}, \ and\ \bibinfo {author} {\bibfnamefont
  {A.}~\bibnamefont {Hoffmann}},\ }\href {\doibase
  10.1103/physrevapplied.8.064026} {\bibfield  {journal} {\bibinfo  {journal}
  {Physical Review Applied}\ }\textbf {\bibinfo {volume} {8}},\ \bibinfo
  {pages} {064026} (\bibinfo {year} {2017})}\BibitemShut {NoStop}%
\bibitem [{\citenamefont {Bingham}\ \emph {et~al.}(2021)\citenamefont
  {Bingham}, \citenamefont {Rooke}, \citenamefont {Park}, \citenamefont
  {Simon}, \citenamefont {Zhu}, \citenamefont {Zhang}, \citenamefont {Batley},
  \citenamefont {Watts}, \citenamefont {Leighton}, \citenamefont {Dahmen},\
  and\ \citenamefont {Schiffer}}]{bingham2021experimental}%
  \BibitemOpen
  \bibfield  {author} {\bibinfo {author} {\bibfnamefont {N.~S.}\ \bibnamefont
  {Bingham}}, \bibinfo {author} {\bibfnamefont {S.}~\bibnamefont {Rooke}},
  \bibinfo {author} {\bibfnamefont {J.}~\bibnamefont {Park}}, \bibinfo {author}
  {\bibfnamefont {A.}~\bibnamefont {Simon}}, \bibinfo {author} {\bibfnamefont
  {W.}~\bibnamefont {Zhu}}, \bibinfo {author} {\bibfnamefont {X.}~\bibnamefont
  {Zhang}}, \bibinfo {author} {\bibfnamefont {J.}~\bibnamefont {Batley}},
  \bibinfo {author} {\bibfnamefont {J.~D.}\ \bibnamefont {Watts}}, \bibinfo
  {author} {\bibfnamefont {C.}~\bibnamefont {Leighton}}, \bibinfo {author}
  {\bibfnamefont {K.~A.}\ \bibnamefont {Dahmen}}, \ and\ \bibinfo {author}
  {\bibfnamefont {P.}~\bibnamefont {Schiffer}},\ }\href {\doibase
  10.1103/physrevlett.127.207203} {\bibfield  {journal} {\bibinfo  {journal}
  {Physical Review Letters}\ }\textbf {\bibinfo {volume} {127}},\ \bibinfo
  {pages} {207203} (\bibinfo {year} {2021})}\BibitemShut {NoStop}%
\bibitem [{\citenamefont {Arnalds}\ \emph {et~al.}(2012)\citenamefont
  {Arnalds}, \citenamefont {Farhan}, \citenamefont {Chopdekar}, \citenamefont
  {Kapaklis}, \citenamefont {Balan}, \citenamefont {Papaioannou}, \citenamefont
  {Ahlberg}, \citenamefont {Nolting}, \citenamefont {Heyderman},\ and\
  \citenamefont {Hjörvarsson}}]{Arnalds_2012_APL}%
  \BibitemOpen
  \bibfield  {author} {\bibinfo {author} {\bibfnamefont {U.~B.}\ \bibnamefont
  {Arnalds}}, \bibinfo {author} {\bibfnamefont {A.}~\bibnamefont {Farhan}},
  \bibinfo {author} {\bibfnamefont {R.~V.}\ \bibnamefont {Chopdekar}}, \bibinfo
  {author} {\bibfnamefont {V.}~\bibnamefont {Kapaklis}}, \bibinfo {author}
  {\bibfnamefont {A.}~\bibnamefont {Balan}}, \bibinfo {author} {\bibfnamefont
  {E.~T.}\ \bibnamefont {Papaioannou}}, \bibinfo {author} {\bibfnamefont
  {M.}~\bibnamefont {Ahlberg}}, \bibinfo {author} {\bibfnamefont
  {F.}~\bibnamefont {Nolting}}, \bibinfo {author} {\bibfnamefont {L.~J.}\
  \bibnamefont {Heyderman}}, \ and\ \bibinfo {author} {\bibfnamefont
  {B.}~\bibnamefont {Hjörvarsson}},\ }\href {\doibase 10.1063/1.4751844}
  {\bibfield  {journal} {\bibinfo  {journal} {Applied Physics Letters}\
  }\textbf {\bibinfo {volume} {101}},\ \bibinfo {pages} {112404} (\bibinfo
  {year} {2012})}\BibitemShut {NoStop}%
\bibitem [{\citenamefont {Andersson}\ \emph {et~al.}(2016)\citenamefont
  {Andersson}, \citenamefont {Pappas}, \citenamefont {Stopfel}, \citenamefont
  {Östman}, \citenamefont {Stein}, \citenamefont {Nordblad}, \citenamefont
  {Mathieu}, \citenamefont {Hjörvarsson},\ and\ \citenamefont
  {Kapaklis}}]{Andersson_2016_SciRep}%
  \BibitemOpen
  \bibfield  {author} {\bibinfo {author} {\bibfnamefont {M.~S.}\ \bibnamefont
  {Andersson}}, \bibinfo {author} {\bibfnamefont {S.~D.}\ \bibnamefont
  {Pappas}}, \bibinfo {author} {\bibfnamefont {H.}~\bibnamefont {Stopfel}},
  \bibinfo {author} {\bibfnamefont {E.}~\bibnamefont {Östman}}, \bibinfo
  {author} {\bibfnamefont {A.}~\bibnamefont {Stein}}, \bibinfo {author}
  {\bibfnamefont {P.}~\bibnamefont {Nordblad}}, \bibinfo {author}
  {\bibfnamefont {R.}~\bibnamefont {Mathieu}}, \bibinfo {author} {\bibfnamefont
  {B.}~\bibnamefont {Hjörvarsson}}, \ and\ \bibinfo {author} {\bibfnamefont
  {V.}~\bibnamefont {Kapaklis}},\ }\href {\doibase 10.1038/srep37097}
  {\bibfield  {journal} {\bibinfo  {journal} {Scientific Reports}\ }\textbf
  {\bibinfo {volume} {6}},\ \bibinfo {pages} {37097} (\bibinfo {year}
  {2016})}\BibitemShut {NoStop}%
\bibitem [{\citenamefont {Pohlit}\ \emph {et~al.}(2020)\citenamefont {Pohlit},
  \citenamefont {Muscas}, \citenamefont {Chioar}, \citenamefont {Stopfel},
  \citenamefont {Ciuciulkaite}, \citenamefont {{\"O}stman}, \citenamefont
  {Pappas}, \citenamefont {Stein}, \citenamefont {Hj{\"o}rvarsson},
  \citenamefont {J{\"o}nsson},\ and\ \citenamefont
  {Kapaklis}}]{Pohlit_PRB_2020}%
  \BibitemOpen
  \bibfield  {author} {\bibinfo {author} {\bibfnamefont {M.}~\bibnamefont
  {Pohlit}}, \bibinfo {author} {\bibfnamefont {G.}~\bibnamefont {Muscas}},
  \bibinfo {author} {\bibfnamefont {I.-A.}\ \bibnamefont {Chioar}}, \bibinfo
  {author} {\bibfnamefont {H.}~\bibnamefont {Stopfel}}, \bibinfo {author}
  {\bibfnamefont {A.}~\bibnamefont {Ciuciulkaite}}, \bibinfo {author}
  {\bibfnamefont {E.}~\bibnamefont {{\"O}stman}}, \bibinfo {author}
  {\bibfnamefont {S.~D.}\ \bibnamefont {Pappas}}, \bibinfo {author}
  {\bibfnamefont {A.}~\bibnamefont {Stein}}, \bibinfo {author} {\bibfnamefont
  {B.}~\bibnamefont {Hj{\"o}rvarsson}}, \bibinfo {author} {\bibfnamefont
  {P.~E.}\ \bibnamefont {J{\"o}nsson}}, \ and\ \bibinfo {author} {\bibfnamefont
  {V.}~\bibnamefont {Kapaklis}},\ }\href {\doibase 10.1103/PhysRevB.101.134404}
  {\bibfield  {journal} {\bibinfo  {journal} {Physical Review B}\ }\textbf
  {\bibinfo {volume} {101}},\ \bibinfo {pages} {134404} (\bibinfo {year}
  {2020})}\BibitemShut {NoStop}%
\bibitem [{\citenamefont {{\"O}stman}\ \emph
  {et~al.}(2018{\natexlab{a}})\citenamefont {{\"O}stman}, \citenamefont
  {Arnalds}, \citenamefont {Kapaklis}, \citenamefont {Taroni},\ and\
  \citenamefont {Hj{\"o}rvarsson}}]{ostman_ising-like_2018}%
  \BibitemOpen
  \bibfield  {author} {\bibinfo {author} {\bibfnamefont {E.}~\bibnamefont
  {{\"O}stman}}, \bibinfo {author} {\bibfnamefont {U.~B.}\ \bibnamefont
  {Arnalds}}, \bibinfo {author} {\bibfnamefont {V.}~\bibnamefont {Kapaklis}},
  \bibinfo {author} {\bibfnamefont {A.}~\bibnamefont {Taroni}}, \ and\ \bibinfo
  {author} {\bibfnamefont {B.}~\bibnamefont {Hj{\"o}rvarsson}},\ }\href
  {\doibase 10.1088/1361-648X/aad0c1} {\bibfield  {journal} {\bibinfo
  {journal} {Journal of Physics: Condensed Matter}\ }\textbf {\bibinfo {volume}
  {30}},\ \bibinfo {pages} {365301} (\bibinfo {year}
  {2018}{\natexlab{a}})}\BibitemShut {NoStop}%
\bibitem [{\citenamefont {Arnalds}\ \emph {et~al.}(2016)\citenamefont
  {Arnalds}, \citenamefont {Chico}, \citenamefont {Stopfel}, \citenamefont
  {Kapaklis}, \citenamefont {B{\"a}renbold}, \citenamefont {Verschuuren},
  \citenamefont {Wolff}, \citenamefont {Neu}, \citenamefont {Bergman},\ and\
  \citenamefont {Hj{\"o}rvarsson}}]{arnalds2016new}%
  \BibitemOpen
  \bibfield  {author} {\bibinfo {author} {\bibfnamefont {U.~B.}\ \bibnamefont
  {Arnalds}}, \bibinfo {author} {\bibfnamefont {J.}~\bibnamefont {Chico}},
  \bibinfo {author} {\bibfnamefont {H.}~\bibnamefont {Stopfel}}, \bibinfo
  {author} {\bibfnamefont {V.}~\bibnamefont {Kapaklis}}, \bibinfo {author}
  {\bibfnamefont {O.}~\bibnamefont {B{\"a}renbold}}, \bibinfo {author}
  {\bibfnamefont {M.~A.}\ \bibnamefont {Verschuuren}}, \bibinfo {author}
  {\bibfnamefont {U.}~\bibnamefont {Wolff}}, \bibinfo {author} {\bibfnamefont
  {V.}~\bibnamefont {Neu}}, \bibinfo {author} {\bibfnamefont {A.}~\bibnamefont
  {Bergman}}, \ and\ \bibinfo {author} {\bibfnamefont {B.}~\bibnamefont
  {Hj{\"o}rvarsson}},\ }\href {\doibase 10.1088/1367-2630/18/2/023008}
  {\bibfield  {journal} {\bibinfo  {journal} {New Journal of Physics}\ }\textbf
  {\bibinfo {volume} {18}},\ \bibinfo {pages} {023008} (\bibinfo {year}
  {2016})}\BibitemShut {NoStop}%
\bibitem [{\citenamefont {Arnalds}\ \emph {et~al.}(2014)\citenamefont
  {Arnalds}, \citenamefont {Ahlberg}, \citenamefont {Brewer}, \citenamefont
  {Kapaklis}, \citenamefont {Papaioannou}, \citenamefont {Karimipour},
  \citenamefont {Korelis}, \citenamefont {Stein}, \citenamefont {{\'O}lafsson},
  \citenamefont {Hase},\ and\ \citenamefont {Hj{\"o}rvarsson}}]{Arnalds_XY}%
  \BibitemOpen
  \bibfield  {author} {\bibinfo {author} {\bibfnamefont {U.~B.}\ \bibnamefont
  {Arnalds}}, \bibinfo {author} {\bibfnamefont {M.}~\bibnamefont {Ahlberg}},
  \bibinfo {author} {\bibfnamefont {M.~S.}\ \bibnamefont {Brewer}}, \bibinfo
  {author} {\bibfnamefont {V.}~\bibnamefont {Kapaklis}}, \bibinfo {author}
  {\bibfnamefont {E.~T.}\ \bibnamefont {Papaioannou}}, \bibinfo {author}
  {\bibfnamefont {M.}~\bibnamefont {Karimipour}}, \bibinfo {author}
  {\bibfnamefont {P.}~\bibnamefont {Korelis}}, \bibinfo {author} {\bibfnamefont
  {A.}~\bibnamefont {Stein}}, \bibinfo {author} {\bibfnamefont
  {S.}~\bibnamefont {{\'O}lafsson}}, \bibinfo {author} {\bibfnamefont
  {T.~P.~A.}\ \bibnamefont {Hase}}, \ and\ \bibinfo {author} {\bibfnamefont
  {B.}~\bibnamefont {Hj{\"o}rvarsson}},\ }\href {\doibase 10.1063/1.4891479}
  {\bibfield  {journal} {\bibinfo  {journal} {Applied Physics Letters}\
  }\textbf {\bibinfo {volume} {105}},\ \bibinfo {pages} {042409} (\bibinfo
  {year} {2014})}\BibitemShut {NoStop}%
\bibitem [{\citenamefont {Skovdal}\ \emph {et~al.}(2021)\citenamefont
  {Skovdal}, \citenamefont {Strandqvist}, \citenamefont {Stopfel},
  \citenamefont {Pohlit}, \citenamefont {Warnatz}, \citenamefont
  {Sl{\"o}etjes}, \citenamefont {Kapaklis},\ and\ \citenamefont
  {Hj{\"o}rvarsson}}]{skovdal2021temperature}%
  \BibitemOpen
  \bibfield  {author} {\bibinfo {author} {\bibfnamefont {B.~E.}\ \bibnamefont
  {Skovdal}}, \bibinfo {author} {\bibfnamefont {N.}~\bibnamefont
  {Strandqvist}}, \bibinfo {author} {\bibfnamefont {H.}~\bibnamefont
  {Stopfel}}, \bibinfo {author} {\bibfnamefont {M.}~\bibnamefont {Pohlit}},
  \bibinfo {author} {\bibfnamefont {T.}~\bibnamefont {Warnatz}}, \bibinfo
  {author} {\bibfnamefont {S.~D.}\ \bibnamefont {Sl{\"o}etjes}}, \bibinfo
  {author} {\bibfnamefont {V.}~\bibnamefont {Kapaklis}}, \ and\ \bibinfo
  {author} {\bibfnamefont {B.}~\bibnamefont {Hj{\"o}rvarsson}},\ }\href
  {\doibase 10.1103/PhysRevB.104.014434} {\bibfield  {journal} {\bibinfo
  {journal} {Physical Review B}\ }\textbf {\bibinfo {volume} {104}},\ \bibinfo
  {pages} {014434} (\bibinfo {year} {2021})}\BibitemShut {NoStop}%
\bibitem [{\citenamefont {Gliga}\ \emph {et~al.}(2015)\citenamefont {Gliga},
  \citenamefont {K{\'a}kay}, \citenamefont {Heyderman}, \citenamefont
  {Hertel},\ and\ \citenamefont {Heinonen}}]{Gliga_PRB_2015}%
  \BibitemOpen
  \bibfield  {author} {\bibinfo {author} {\bibfnamefont {S.}~\bibnamefont
  {Gliga}}, \bibinfo {author} {\bibfnamefont {A.}~\bibnamefont {K{\'a}kay}},
  \bibinfo {author} {\bibfnamefont {L.~J.}\ \bibnamefont {Heyderman}}, \bibinfo
  {author} {\bibfnamefont {R.}~\bibnamefont {Hertel}}, \ and\ \bibinfo {author}
  {\bibfnamefont {O.~G.}\ \bibnamefont {Heinonen}},\ }\href {\doibase
  10.1103/PhysRevB.92.060413} {\bibfield  {journal} {\bibinfo  {journal}
  {Physical Review B}\ }\textbf {\bibinfo {volume} {92}},\ \bibinfo {pages}
  {060413} (\bibinfo {year} {2015})}\BibitemShut {NoStop}%
\bibitem [{\citenamefont {{Sl{\"o}etjes}}\ \emph {et~al.}(2021)\citenamefont
  {{Sl{\"o}etjes}}, \citenamefont {{Hj{\"o}rvarsson}},\ and\ \citenamefont
  {{Kapaklis}}}]{Sloetjes_arXiv_2020}%
  \BibitemOpen
  \bibfield  {author} {\bibinfo {author} {\bibfnamefont {S.~D.}\ \bibnamefont
  {{Sl{\"o}etjes}}}, \bibinfo {author} {\bibfnamefont {B.}~\bibnamefont
  {{Hj{\"o}rvarsson}}}, \ and\ \bibinfo {author} {\bibfnamefont
  {V.}~\bibnamefont {{Kapaklis}}},\ }\href {\doibase 10.1063/5.0048789}
  {\bibfield  {journal} {\bibinfo  {journal} {Applied Physics Letters}\
  }\textbf {\bibinfo {volume} {118}},\ \bibinfo {pages} {142407} (\bibinfo
  {year} {2021})}\BibitemShut {NoStop}%
\bibitem [{\citenamefont {Shinjo}(2000)}]{shinjo_magnetic_2000}%
  \BibitemOpen
  \bibfield  {author} {\bibinfo {author} {\bibfnamefont {T.}~\bibnamefont
  {Shinjo}},\ }\href {\doibase 10.1126/science.289.5481.930} {\bibfield
  {journal} {\bibinfo  {journal} {Science}\ }\textbf {\bibinfo {volume}
  {289}},\ \bibinfo {pages} {930} (\bibinfo {year} {2000})}\BibitemShut
  {NoStop}%
\bibitem [{\citenamefont {Kl{\"a}ui}\ \emph {et~al.}(2003)\citenamefont
  {Kl{\"a}ui}, \citenamefont {Vaz}, \citenamefont {Lopez-Diaz},\ and\
  \citenamefont {Bland}}]{Klaui_vortx_2003}%
  \BibitemOpen
  \bibfield  {author} {\bibinfo {author} {\bibfnamefont {M.}~\bibnamefont
  {Kl{\"a}ui}}, \bibinfo {author} {\bibfnamefont {C.~A.~F.}\ \bibnamefont
  {Vaz}}, \bibinfo {author} {\bibfnamefont {L.}~\bibnamefont {Lopez-Diaz}}, \
  and\ \bibinfo {author} {\bibfnamefont {J.~A.~C.}\ \bibnamefont {Bland}},\
  }\href {\doibase 10.1088/0953-8984/15/21/201} {\bibfield  {journal} {\bibinfo
   {journal} {Journal of Physics: Condensed Matter}\ }\textbf {\bibinfo
  {volume} {15}},\ \bibinfo {pages} {R985} (\bibinfo {year}
  {2003})}\BibitemShut {NoStop}%
\bibitem [{\citenamefont {{Skovdal}}\ \emph {et~al.}(2022)\citenamefont
  {{Skovdal}}, \citenamefont {{P{\'a}lsson}}, \citenamefont {{Holdsworth}},\
  and\ \citenamefont {{Hj{\"o}rvarsson}}}]{tricritical}%
  \BibitemOpen
  \bibfield  {author} {\bibinfo {author} {\bibfnamefont {B.~E.}\ \bibnamefont
  {{Skovdal}}}, \bibinfo {author} {\bibfnamefont {G.~K.}\ \bibnamefont
  {{P{\'a}lsson}}}, \bibinfo {author} {\bibfnamefont {P.~C.~W.}\ \bibnamefont
  {{Holdsworth}}}, \ and\ \bibinfo {author} {\bibfnamefont {B.}~\bibnamefont
  {{Hj{\"o}rvarsson}}},\ }\href@noop {} {\bibfield  {journal} {\bibinfo
  {journal} {arXiv e-prints}\ ,\ \bibinfo {eid} {arXiv:2204.10065}} (\bibinfo
  {year} {2022})},\ \Eprint {http://arxiv.org/abs/2204.10065} {arXiv:2204.10065
  [cond-mat.mes-hall]} \BibitemShut {NoStop}%
\bibitem [{\citenamefont {Phatak}\ \emph {et~al.}(2011)\citenamefont {Phatak},
  \citenamefont {Petford-Long}, \citenamefont {Heinonen}, \citenamefont
  {Tanase},\ and\ \citenamefont {De~Graef}}]{phatak2011nanoscale}%
  \BibitemOpen
  \bibfield  {author} {\bibinfo {author} {\bibfnamefont {C.}~\bibnamefont
  {Phatak}}, \bibinfo {author} {\bibfnamefont {A.}~\bibnamefont
  {Petford-Long}}, \bibinfo {author} {\bibfnamefont {O.}~\bibnamefont
  {Heinonen}}, \bibinfo {author} {\bibfnamefont {M.}~\bibnamefont {Tanase}}, \
  and\ \bibinfo {author} {\bibfnamefont {M.}~\bibnamefont {De~Graef}},\ }\href
  {\doibase 10.1103/physrevb.83.174431} {\bibfield  {journal} {\bibinfo
  {journal} {Physical Review B}\ }\textbf {\bibinfo {volume} {83}},\ \bibinfo
  {pages} {174431} (\bibinfo {year} {2011})}\BibitemShut {NoStop}%
\bibitem [{\citenamefont {Leliaert}\ \emph {et~al.}(2017)\citenamefont
  {Leliaert}, \citenamefont {Mulkers}, \citenamefont {De~Clercq}, \citenamefont
  {Coene}, \citenamefont {Dvornik},\ and\ \citenamefont
  {Van~Waeyenberge}}]{leliaert2017adaptively}%
  \BibitemOpen
  \bibfield  {author} {\bibinfo {author} {\bibfnamefont {J.}~\bibnamefont
  {Leliaert}}, \bibinfo {author} {\bibfnamefont {J.}~\bibnamefont {Mulkers}},
  \bibinfo {author} {\bibfnamefont {J.}~\bibnamefont {De~Clercq}}, \bibinfo
  {author} {\bibfnamefont {A.}~\bibnamefont {Coene}}, \bibinfo {author}
  {\bibfnamefont {M.}~\bibnamefont {Dvornik}}, \ and\ \bibinfo {author}
  {\bibfnamefont {B.}~\bibnamefont {Van~Waeyenberge}},\ }\href {\doibase
  10.1063/1.5003957} {\bibfield  {journal} {\bibinfo  {journal} {AIP Advances}\
  }\textbf {\bibinfo {volume} {7}},\ \bibinfo {pages} {125010} (\bibinfo {year}
  {2017})}\BibitemShut {NoStop}%
\bibitem [{\citenamefont {{\"O}stman}\ \emph
  {et~al.}(2018{\natexlab{b}})\citenamefont {{\"O}stman}, \citenamefont
  {Stopfel}, \citenamefont {Chioar}, \citenamefont {Arnalds}, \citenamefont
  {Stein}, \citenamefont {Kapaklis},\ and\ \citenamefont
  {Hj{\"o}rvarsson}}]{ostman_interaction_2018}%
  \BibitemOpen
  \bibfield  {author} {\bibinfo {author} {\bibfnamefont {E.}~\bibnamefont
  {{\"O}stman}}, \bibinfo {author} {\bibfnamefont {H.}~\bibnamefont {Stopfel}},
  \bibinfo {author} {\bibfnamefont {I.-A.}\ \bibnamefont {Chioar}}, \bibinfo
  {author} {\bibfnamefont {U.~B.}\ \bibnamefont {Arnalds}}, \bibinfo {author}
  {\bibfnamefont {A.}~\bibnamefont {Stein}}, \bibinfo {author} {\bibfnamefont
  {V.}~\bibnamefont {Kapaklis}}, \ and\ \bibinfo {author} {\bibfnamefont
  {B.}~\bibnamefont {Hj{\"o}rvarsson}},\ }\href {\doibase
  10.1038/s41567-017-0027-2} {\bibfield  {journal} {\bibinfo  {journal} {Nature
  Physics}\ }\textbf {\bibinfo {volume} {14}},\ \bibinfo {pages} {375}
  (\bibinfo {year} {2018}{\natexlab{b}})}\BibitemShut {NoStop}%
\bibitem [{\citenamefont {P{\"a}rnaste}\ \emph {et~al.}(2007)\citenamefont
  {P{\"a}rnaste}, \citenamefont {Marcellini}, \citenamefont {Holmstr{\"o}m},
  \citenamefont {Bock}, \citenamefont {Fransson}, \citenamefont {Eriksson},\
  and\ \citenamefont {Hj{\"o}rvarsson}}]{parnaste2007dimensionality}%
  \BibitemOpen
  \bibfield  {author} {\bibinfo {author} {\bibfnamefont {M.}~\bibnamefont
  {P{\"a}rnaste}}, \bibinfo {author} {\bibfnamefont {M.}~\bibnamefont
  {Marcellini}}, \bibinfo {author} {\bibfnamefont {E.}~\bibnamefont
  {Holmstr{\"o}m}}, \bibinfo {author} {\bibfnamefont {N.}~\bibnamefont {Bock}},
  \bibinfo {author} {\bibfnamefont {J.}~\bibnamefont {Fransson}}, \bibinfo
  {author} {\bibfnamefont {O.}~\bibnamefont {Eriksson}}, \ and\ \bibinfo
  {author} {\bibfnamefont {B.}~\bibnamefont {Hj{\"o}rvarsson}},\ }\href
  {\doibase 10.1088/0953-8984/19/24/246213} {\bibfield  {journal} {\bibinfo
  {journal} {Journal of physics: Condensed matter}\ }\textbf {\bibinfo {volume}
  {19}},\ \bibinfo {pages} {246213} (\bibinfo {year} {2007})}\BibitemShut
  {NoStop}%
\bibitem [{\citenamefont {Vansteenkiste}\ \emph {et~al.}(2014)\citenamefont
  {Vansteenkiste}, \citenamefont {Leliaert}, \citenamefont {Dvornik},
  \citenamefont {Helsen}, \citenamefont {Garcia-Sanchez},\ and\ \citenamefont
  {Van~Waeyenberge}}]{vansteenkiste_design_2014}%
  \BibitemOpen
  \bibfield  {author} {\bibinfo {author} {\bibfnamefont {A.}~\bibnamefont
  {Vansteenkiste}}, \bibinfo {author} {\bibfnamefont {J.}~\bibnamefont
  {Leliaert}}, \bibinfo {author} {\bibfnamefont {M.}~\bibnamefont {Dvornik}},
  \bibinfo {author} {\bibfnamefont {M.}~\bibnamefont {Helsen}}, \bibinfo
  {author} {\bibfnamefont {F.}~\bibnamefont {Garcia-Sanchez}}, \ and\ \bibinfo
  {author} {\bibfnamefont {B.}~\bibnamefont {Van~Waeyenberge}},\ }\href
  {\doibase 10.1063/1.4899186} {\bibfield  {journal} {\bibinfo  {journal} {AIP
  Advances}\ }\textbf {\bibinfo {volume} {4}},\ \bibinfo {pages} {107133}
  (\bibinfo {year} {2014})}\BibitemShut {NoStop}%
\end{thebibliography}

%

\end{document}